\documentclass{jfm}
\usepackage{graphicx}
\usepackage{epstopdf, epsfig}
\usepackage[thinc]{esdiff}
\usepackage{mathtools}
\usepackage{graphicx}
\usepackage{dcolumn}
\usepackage{bm}
\usepackage{soul}
\usepackage{color}
\usepackage{textcomp}
\usepackage{algorithmic}
\usepackage{array}
\usepackage{lipsum}
\usepackage{hyperref}
\hypersetup{
    colorlinks=true,
    linkcolor = black,
    anchorcolor = blue,
    citecolor = blue,
    filecolor = blue,
    urlcolor = blue
}
\usepackage{textcomp}
\usepackage{dirtytalk}
\usepackage{soul}
\usepackage{comment}
\usepackage{tabularx}
\usepackage{caption}
\usepackage{subcaption}
\usepackage{amssymb}
\usepackage{amsmath} 
\usepackage{textcomp}
\usepackage{breqn}
\usepackage{float}

\newcommand{\Vol}{\rotatebox[origin=c]{180}{\ensuremath{A}}}

\shorttitle{Spreading dynamics of droplets impacting on oscillating hydrophobic substrates}
\shortauthor{A. Potnis and A. Saha}

\title{Spreading dynamics of droplets impacting on oscillating hydrophobic substrates}

\author{Aditya Potnis\aff{1},
 \and Abhishek Saha\aff{1}
 \corresp{\email{asaha@eng.ucsd.edu}}}

\affiliation{\aff{1}Department of Mechanical and Aerospace Engineering, University of California San Diego, 
La Jolla, CA 92093, USA}

\begin{document}

\maketitle

\begin{abstract}
Droplet impact on oscillating substrates is important for both natural and industrial processes. Recognizing the importance of the dynamics that arise from the interplay between droplet transport and substrate motion, in this work, we present an experimental investigation of the spreading of a droplet impacting a sinusoidally oscillating hydrophobic substrate. We particularly focus on the maximum spread of droplets as a function of various parameters of substrate oscillation. 
We first quantify the maximum spreading diameter attained by the droplets as a function of frequency, amplitude of vibration, and phase at the impact for various impact velocities. We highlight that there can be two stages of spreading. \textit{Stage-I}, which is observed at all impact conditions, is controlled by the droplet inertia and affected by the substrate oscillation. For certain conditions, a \textit{Stage-II} spreading is also observed, which occurs during the retraction process of \textit{Stage-I} due to additional energies imparted by the substrate oscillation. Subsequently, we derive scaling analyses to predict the maximum spreading diameters and the time for this maximum spread for both \textit{Stage-I} and \textit{Stage-II}. Furthermore, we identify the necessary condition for \textit{Stage-II} spreading to be greater than the \textit{Stage-I}. The results will enable optimization of the parameters in applications where substrate oscillation is used to control the droplet spread and, thus, heat and mass transfer between the droplet and the substrate.    

\end{abstract}


\section{Introduction}
\label{sec:intro}

Droplet impacts on stationary and non-stationary surfaces are frequently witnessed in our daily lives. For instance, raindrops falling onto leaves \citep{gart2015droplet,park2020dynamics} and onto beating wings of birds \citep{zhang2019non} and insects entail fluid-structure interaction wherein movement of the surfaces greatly affects the outcomes of such impacts and thus governs numerous phenomenon such as the spread of pathogens, environmental aerosol dispersion, and repulsion of liquid from natural surfaces. Understanding this droplet-surface interaction is necessary to improve agricultural practices and to design superior bio-inspired water-repellent media. In addition, droplet impact on vibrating surfaces is a critical aspect of numerous industrial processes. Controlling the deposition of functional droplets can be useful in sprays for thermal coating \citep{tropea2000modeling, saha-SCT-2009b}. The impact of droplets on substrates is also critical for applications such as inkjet printing and in additive manufacturing 
\citep{martin2008inkjet, tang-JFM-2019, lohse2022ARFM}, where the post-impact spreading of droplet determines the quality of the final product. The spreading dynamics of impacted droplets also control the efficacy of spray cooling (or heating) of surfaces where the heat transfer is proportional to the contact area and contact time between the droplet and the substrate \citep{breitenbach2018drop}. Furthermore, the spreading is an important parameter in applications where the impact promotes chemical reactions during surface treatments or mass transfers during cleaning processes. Owing to its broad applications, droplet impact on stationary and non-stationary substrates is of scientific interest. Hence, a large volume of studies investigated the mechanistic description of the dynamics, and various post-impact outcomes, which will be reviewed next.

Over the past century, extensive work has been carried out to investigate the impact of different liquids on stationary substrates with a variety of surface properties and morphologies. Early studies published by \citet{worthington1877iii,worthington1877xxviii} explored the impact of water and mercury on a static glass surface. Since then, numerous studies have explored the various aspects of droplet impact on stationary media, which were periodically summarized in reviews by  \citet{yarin2006drop,khojasteh2016droplet,josserand2016drop}. 
These studies have established that the outcome of impact and post-impact dynamics are governed by the properties of both liquid and substrate. The phenomena observed during impact are the results of the balance between forces involving impact inertia, capillary force (surface tension), viscous dissipation, and gravitational force. The relative strength of these forces can be quantified using non-dimensional numbers, such as Weber number ($We$), Reynolds number ($Re$), and Froude number ($Fr$). 
\begin{equation}
We = \frac{\rho V_0^2 D_0}{\gamma},\quad Re = \frac{\rho V_0 D_0}{\mu},\quad Fr = \frac{V_0^2}{g D_0}
\label{eq: non-dim numbers}
\end{equation}
where, $\rho$ is the liquid density, $V_0$ is the impact velocity of droplet, $D_0$ is the initial droplet diameter, $\gamma$ is the air-liquid surface tension, $\mu$ is the dynamic viscosity of the liquid, $g$ is the gravitational acceleration. 
Previous studies have found distinct characteristics of impact differentiated by their governing mechanisms and classified as either viscous regime (low $We$ and $Re$) or inertia-capillary regime (high $We$ and $Re$). \citet{rioboo2001outcomes} presented the different qualitative outcomes of droplet impact on solid surfaces characterized, namely as prompt splash, corona splash, rebound, partial rebound, deposition, and receding breakup. Splashing generally occurs when the inertia of the droplet overcomes surface tension during high $We$ impacts \citep{PhysRevLett.94.184505,liu2010splashing,hao2017splash,mandre2012mechanism,khabakhpasheva2020splashing}. 
For low $We$ impacts, the primary focus of a group of studies was dynamics of droplet deformation and retraction during impacts on various impact surfaces \citep{tang-JFM-2019, bird2009inclined, clanet2004maximal}.
The outcome of droplet impact is generally quantified in the form of geometric parameters, such as spreading factor ($D^* = D/D_0$), defined by the ratio of instantaneous droplet diameter ($D$) to initial diameter and the instantaneous height of the deformed droplet. Some studies 
\citep{antonini2013drop, bartolo2005retraction} also analyzed key timescales, including contact time, rebounding time, etc. 
The time evolution of the spreading factor was investigated in several investigations, which highlighted the maximum spread factor ($D^*_{max}$). Experimentally it was shown that the evolution of $D^*$, and $D^*_{max}$ depend on $We$, $Re$ and substrate properties including, wettability (substrate contact angle), surface roughness, etc~\citep{lagubeau2012spreading,eggers2010drop,ukiwe2005maximum}. The experimental observations were complimented by several theoretical and analytical studies~\citep{pasandideh1996capillary, du2021analytical, chandra1991collision,clanet2004maximal, fedorchenko2005effect}. These approaches, often based on the assumption of simplified geometry of deformed droplets, assess the role of viscous loss along the boundary layer to derive either scaling laws or estimated droplet diameter~\citep{bennett1993splat,pasandideh1996capillary,attane2007energy}. To obtain more detailed insights and to analyze localized dynamics, a large number of studies used numerical simulations \citep{vsikalo2005dynamic,raman2016lattice,tropea2000modeling,rioboo2002time,eggers2010drop,renardy2003pyramidal,wildeman2016spreading}. 

The impact of a droplet on moving surfaces differs from the static surface due to the modification of the relative velocity between the droplet and the substrate and the associated change in the interfacial dynamics. \citet{lee2004control} experimentally studied the impact of low-viscosity droplets on a moving substrate and explored the influence of various trajectories of vertical motion of a surface on the post-impact characteristics of spreading and rebound. They showed that the observed modification in the dynamics could not be solely attributed to the change of relative droplet velocity and that different trajectories of substrate motion with similar relative velocity can cause various degrees of deviation in the dynamics. \citet{weisensee2017droplet} carried out experiments with vertically oscillating rigid as well as elastic surfaces to determine the effect of surface motion on the rebound characteristics of impacting droplets. They provided evidence that the contact time is a strong function of the timescale of oscillation and demonstrated that the phase of surface oscillation at the impact plays a critical role. This was subsequently confirmed by \citet{kim2018dynamics}, who worked with flexible superhydrophobic surfaces with varying natural frequencies. Recent experimental and theoretical work by \citet{upadhyay2021bouncing} on flexible superhydrophobic surfaces showed that a spring-mass system model could estimate the contact time of droplets before the rebound. \citet{moradi2020numerical} used axisymmetric Lattice Boltzmann simulations and confirmed that, for low adhesion surfaces, the amplitude, frequency of oscillation,
and the phase at impact dictate the spreading and rebound characteristics. Similar results were also reported for impact on a superhydrophobic surface using a discrete particle method based on many-body dissipative particle dynamics (MDPD) \citep{lin2022dynamic} and a coupled level-set and volume of fluid (CLSVOF) method \citep{li2022hydrodynamic}. 
Along with reaffirming that oscillation parameters are critical to post-impact droplet dynamics, these studies have also reported that energy dissipation plays a key role in determining droplet dimensions, such as the maximum spreading diameter and height during the spreading process. For high-energy destabilizing impacts, \citet{khabakhpasheva2020splashing} used asymptotic analysis to derive a model and conjectured that splashing is possible for impacts on a rigid vibrating surface but not on elastically supported substrates. While most of these studies employed substrates oscillating parallel to the direction of the impact, some studies investigated the effects of substrate motion perpendicular to the direction of the impact. In general, it is shown that such motions also alter the post-impact behavior and that by controlling the parameters of oscillation, droplet rebound can be suppressed \citep{raman2016rebound, raman2019normal}. 

The above short review of literature has shown that droplet impact dynamics are markedly different for impacts on non-stationary substrates. Although these works have investigated the impact on oscillating surfaces, a vast majority have utilized superhydrophobic surfaces, wherein the droplet is prone to rebound upon impact due to its low affinity to the substrate. Additionally, there is a lack of systematic experimental studies to explore the effect of a wide range of oscillation parameters on the maximum spread of impacting droplets. Our work endeavors to experimentally show that the dynamics of droplet impact can be manipulated by using a vertically oscillating rigid hydrophobic substrate. We investigate the spreading behavior in the `deposition' regime of impact as described in \citet{rioboo2001outcomes}. We will provide evidence that the reported post-impact normalized maximum droplet diameter ($D^*_{max}$) during spreading and the time taken to achieve this quantity ($t_{max}$) is strongly influenced by parameters of surface oscillations namely, amplitude ($A$) and frequency ($f$) of oscillation and phase at impact ($\phi$). Finally, we will provide scaling analyses to theoretically assess the effects of these oscillation parameters on the $t_{max}$ and $D^*_{max}$. 

Next, we will describe the experimental setup. After that, we will present the experimental results of the impact on static and oscillating substrates, although our primary focus will be on the latter. Subsequently, we will present the scaling analyses for theoretical estimations of the maximum spreading of droplets. Finally, we will conclude with a summary of the work.

\section{Experimental setup}
\label{sec:expt_setup}

\begin{figure}
    \centering
    \includegraphics[width=0.8\textwidth]{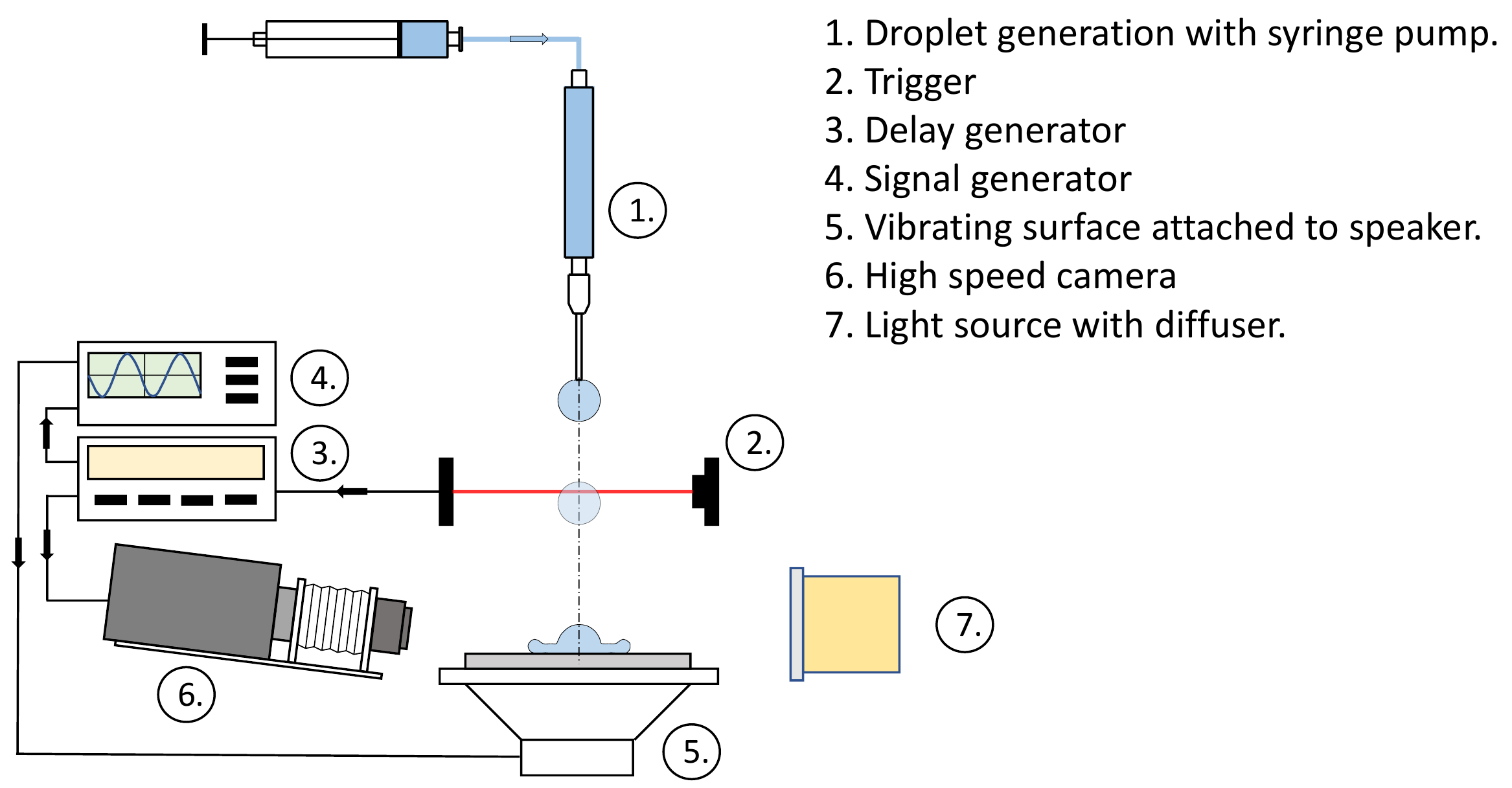}
\caption{Schematic of the experimental setup for droplet impact on a vibrating hydrophobic substrate.}
\label{fig:setup}
\end{figure}

In our experiments, a single droplet of de-ionized ($\textit{DI}$) water ($\rho=998~ kg\textrm{-}m^{-3}$, $\gamma=72~mN\textrm{-}m^{-1}$, and $\mu=0.89~mPa\textrm{-}s$) is generated at the tip of a vertically positioned needle (outer diameter $0.25~mm$) by pushing liquid through it using a syringe pump. When the gravitational force overcomes the surface tension, the droplet detaches, yielding an almost constant initial diameter $\text{D}_0$ ($\approx 2~mm$). Upon detaching from the needle, the droplet takes a near-spherical shape with minimal deformation. The droplet accelerates as it falls downwards and eventually lands on the substrate with an impact velocity $\text{V}_0$. By changing the free-fall distance, $\text{V}_0$ was varied to yield approximate Reynolds number ($Re$, defined in Eq. \ref{eq: non-dim numbers}) and Weber number ($We$, defined in Eq. \ref{eq: non-dim numbers}) in the ranges of $1890<Re<3765$ and $18<We<77$, respectively. Our substrate is a mirrored glass surface mounted on a speaker unit (STAPEZ$\textsuperscript{TM}$ FP-SPK-M glass-composite-diaphragm woofer). The substrate is coated with a commercially available hydrophobic coating (\textit{RainX}), which results in a static contact angle ($\theta_{eq}$) of $\sim 90^{\circ}$ with $\textit{DI}$ water. A schematic of our experimental setup is shown in Figure \ref{fig:setup}. 

A sinusoidal signal from the function generator is used to drive the speaker to provide a controlled substrate oscillation. This results in a vertical oscillation in the substrate in the form $y_{s} = A\ sin(2 \pi f t + \phi)$, where $y_{s}$ is the position of surface measured from the static position, $t$ is time, $A$ is the amplitude of oscillation, $f$ is the frequency, and $\phi$ is the phase at impact. Here, $t=0$ refers to the instance when the droplet impacts the substrate. For this work, we assign $y_s$ to be positive in the upward direction.

A high-speed camera (Phantom V7.3) coupled with a Nikon $50~mm$ lens, 2x teleconverter, and an extension bellow is used to record impact dynamics. The camera and lens are mounted at a slight incline to the substrate to ensure an unrestricted view of the droplet even during high amplitude oscillations (shown in Fig. \ref{sec:expt_setup}). Images are recorded at $14,760$ frames per second with a $512 \times 384 ~pixel^2$ resolution yielding a spatial resolution of approximately $12.9 ~\mu m/pixel$. A high-intensity diffused LED array is used as a backlight. Each experimental condition was repeated at least three times to ensure the repeatability of the result. The ensemble average of the desired quantity is considered, and the standard deviation is recorded to estimate the error. The error associated with the desired phase at impact ($\phi$) is $\pm 0.028\pi$ or $\pm 5^o$.

A laser-based sensor, placed slightly above the impacted surface, was used to detect the proximity of the droplet from the substrate. The signal from the sensor triggers the high-speed camera and the function generator through an external delay generator. The phase of the oscillating substrate at the time of impact (i.e., the phase at impact, $\phi$) is controlled by adding and modulating a delay between the signal from the sensor and the function generator. The frequency and amplitude of the oscillation in the substrate were directly controlled through the function generator (direct digital synthesis, DDS, signal generator from Koolertron), whose output was amplified using an amplifier (Lepai $\textsuperscript{\textregistered}~\text{LP-}220\text{TI}$) before sending it to the speaker to oscillate the substrate. Before the experiments, the relation between the input voltage to the speaker and ensued amplitude of the oscillation was obtained through a detailed calibration process.

High-speed images obtained during the experiments are processed using a custom MATLAB code to extract the quantitative information on droplet spreading. In our study, the instantaneous diameter of the deformed droplet ($D(t)$) is defined as the maximum horizontal extent of the droplet as seen in high-speed images. It is worth noting that this diameter, $D(t)$, is different from the contact diameter (diameter of the contact line), especially during the initial period of the deformation. 

\begin{figure}
    \centering
    \includegraphics[width=1\textwidth]{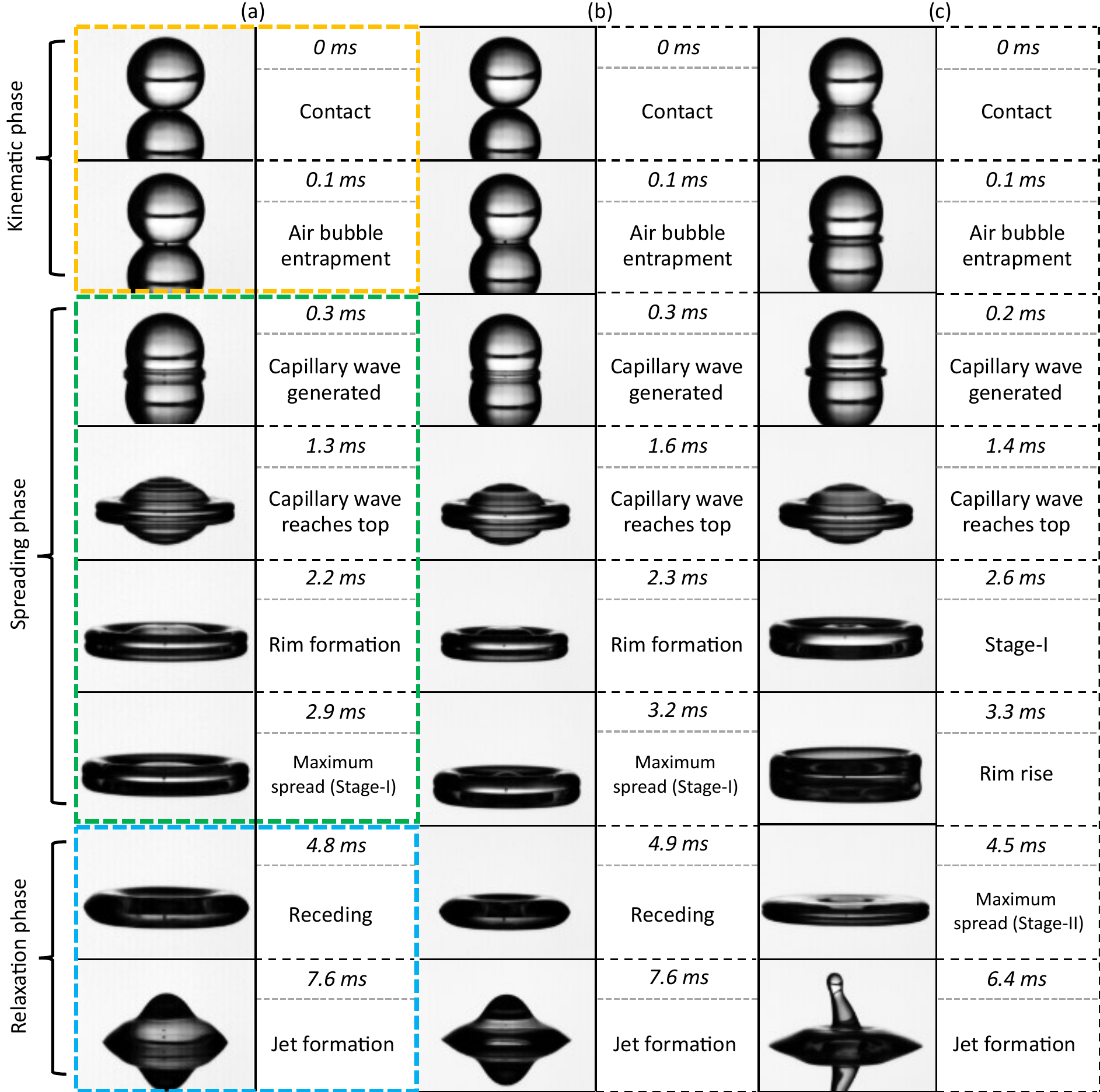}
\caption{high-speed snapshots showing stages of droplet impact for $We = 27$, $Re = 2300$ with (a) static substrate (b) $\text{f}$ = $100\text{Hz}$, $\text{A}$ = $0.25mm$ and $\phi = 3 \pi/4~rad$ (c) $\text{f}$ = $400\text{Hz}$, $\text{A}$ = $0.125mm$ and $\phi = \pi/4~rad$. Multiple peaks in the time evolution of droplet spread are observed for high frequency cases due to the effect of subsequent oscillations as elaborated in Sec. \ref{result_osc}.}
\label{fig:droplet_images}
\end{figure}

\section{Experimental results}
\label{sec:results}

This section will illustrate the experimental findings of our investigation. Our interest lies primarily in the post-impact spreading behavior, particularly the normalized maximal spreading of droplets. Henceforth, in this exposition, we will highlight the observations till the instant of the maximum spread of droplets, with limited scrutiny of the receding phase. After establishing a baseline for our study by outlining the findings of the impact on a static surface, we will present the dynamics for cases with oscillating substrates, which is the focus of this study. The influence of phase at impact ($\phi$), the effect of frequency of oscillation ($f$), along with the effect of oscillation amplitude ($A$) will be addressed systematically. 

\subsection{Impact on static substrates}
\label{result_static}

The speaker was not actuated for these experiments, ensuring the substrate remained static during the impact. The droplet free falls till it contacts the substrate at $t=0$, with the velocity at impact $\text{V}_0$. 
A series of high-speed images for impact on a static surface is shown in Figure. \ref{fig:droplet_images}a. The process begins with the spherical droplet making contact with the stationary substrate. Promptly after the instant of impact ($t< 0.1~ms$), a small air bubble forms near the contact point (also observed for oscillating substrates as shown in Fig. \ref{fig:droplet_images}) due to the entrapment of air between the droplet and surface. This has been observed in previous studies \citep{chandra1991collision,pasandideh1996capillary, bouwhuis2012maximal, tang_pof_2019} 
of droplet impact on both solid and liquid substrates. This bubble formation is caused by non-uniform pressure distribution in the interfacial gas layer trapped between the droplet and impacted interfaces and the eventual collapse of this gas layer at a location away from its center. 
The droplet starts deforming after the impact (Fig. \ref{fig:droplet_images}a), its bottom surface is flattened, and the droplet spreads outwards. This initial stage of spreading (upto $t\approx 0.3~ms$ as illustrated in Figure \ref{fig:droplet_images}a) is referred to as the `kinematic phase' \citep{rioboo2002time} where the impact inertia dominates over capillary and viscous effects in controlling the spreading dynamics.

\begin{figure}
\centering
    \includegraphics[width=1\textwidth]{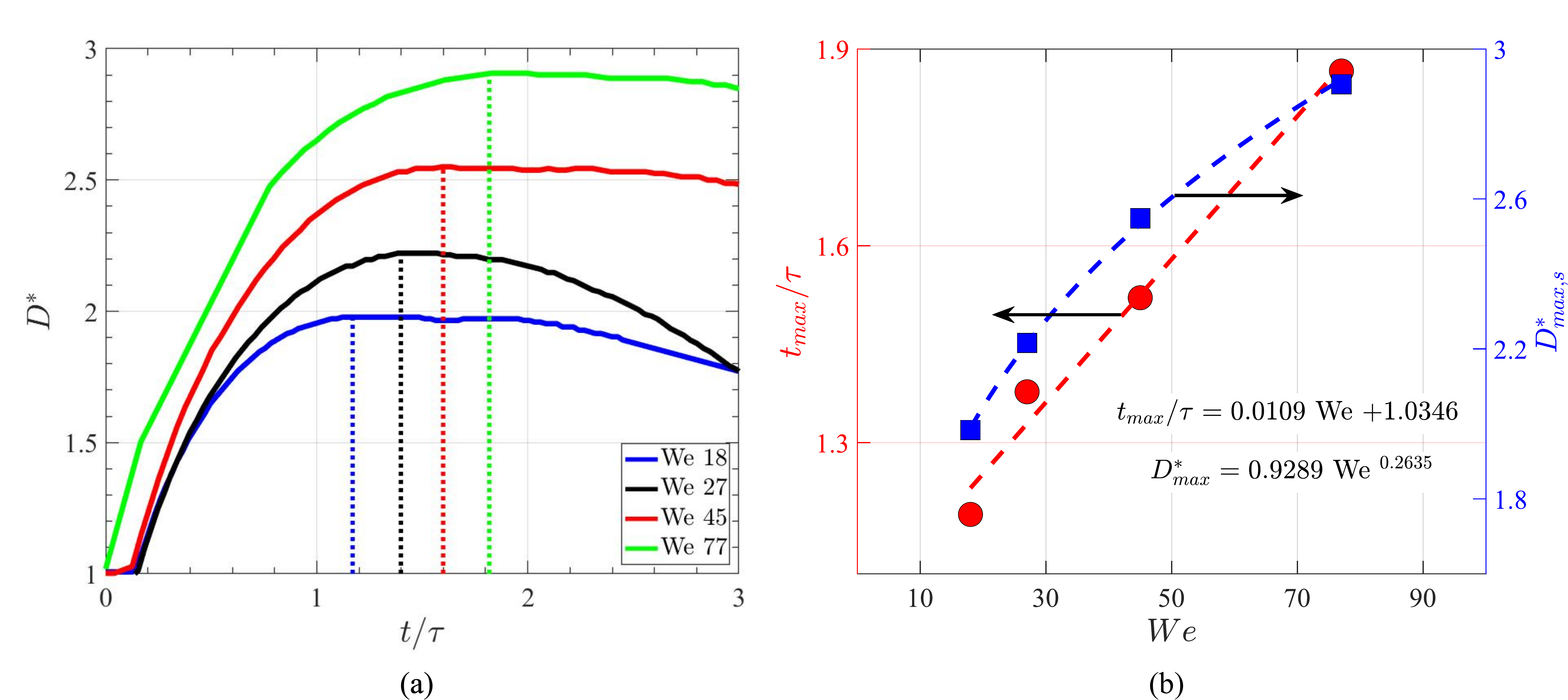}
\caption{(a) Temporal evolution of post-impact normalized droplet diameter ($\text{D}/\text{D}_0$) for static impact for $We = 18, 27, 45$ $\&$ $77$ -- {--} {--} denotes the normalized time $t_{max}/\tau$ for maximum spreading. (b) Normalized maximum spread for static impact ($\text{D}_{max,s}^*$) and normalized time for the maximum spread, with respect to `crashing time'  ($t_{max,s}/\tau$) as functions of $We$ which display a power law, and linear fitting respectively.}
\label{fig:0HzWe_tmaxfit}
\end{figure}

As the deformed portion of the droplet spreads past the initial droplet diameter $D_0$, a lamella is formed, and it rapidly spreads radially, while the upper portion of the droplet remains undeformed, resembling a truncated sphere as seen at $t\approx 0.3~ms$ in Figure \ref{fig:droplet_images}a. It is observed that this undeformed part of the droplet continues to move downward with the velocity (measured at the tip of the droplet) equal to the impact velocity ($\text{V}_0$), an observation also reported in earlier studies \citep{lagubeau2012spreading}. This is the `spreading phase' \citep{rioboo2002time} where surface tension and viscosity begin to affect the spread. At this stage, capillary waves are seen to rise up through the droplet surface, and they travel upwards, eventually reaching the top of the droplet ($\approx 1.3~ms$ in Figure \ref{fig:droplet_images}a), thereby completely deforming it. More details of these waves can be found in studies by \citet{pasandideh1996capillary} and \citet{renardy2003pyramidal}.



As the top of the droplet reaches its lowest point, it no longer looks like a spherical cap but resembles a pancake ($\approx 2.2~ms$ in Figure \ref{fig:droplet_images}a). For inertia-driven impact, the time taken for the droplet to reach this stage is defined as the droplet `crashing time' ($\tau= \text{D}_0/\text{V}_0$) and is the characteristic inertial timescale for such phenomena. At this stage, the droplet has expended most of its kinetic energy. The `spreading phase' continues till the droplet completely deforms and all its kinetic energy is traded for an increase in surface energy, with some energy lost to viscous dissipation. The spreading results in thinning of the lamella and a thick rim is formed on the droplet periphery, creating an almost toroidal geometry (labeled as `maximum spread' at $\approx 2.9~ms$ in Figure \ref{fig:droplet_images}a). For lower $\text{V}_0$ (and $We$) and thus lesser spread, the droplet forms a geometry with a less pronounced rim described as pancake-like rather than toroidal (not shown here). The time history of the instantaneous diameter of the droplet for four different $We$ is shown in Fig. \ref{fig:0HzWe_tmaxfit}a. The normalized maximum diameter ($\text{D}_{max,s}^* = \text{D}_{max,s}/ \text{D}_{0}$), achieved by the droplets, and the normalized time ($\text{t}_{max,s}/\tau$) taken to achieve the maximum spread are shown in Figure \ref{fig:0HzWe_tmaxfit}b. In general, we observe that $\text{t}_{max,s}$ is greater than $\tau$ and the normalized time follows a linear relation ($\text{t}_{max,s}/\tau = 0.0109 We + 1.0346$) with $We$.  
On the other hand, normalized maximum diameter ($\text{D}_{max,s}^*$) displays a power-law dependence on $We$, with an exponent of approximately $1/4$, a behavior also observed in previous studies \citep{clanet2004maximal}. These values of $\text{D}_{max,s}^*$ and $\text{t}_{max,s}$ for static impact are recorded as the `baseline' to juxtapose with oscillating substrate cases and thus, will be used to normalize corresponding length and time scales.

After achieving maximum spread, the system transitions to the `relaxation phase' \citep{rioboo2002time}, and the droplet recedes due to surface tension in an effort to minimize its surface energy. The droplet settles into a damped oscillation of its diameter till viscous losses eventually render it stationary, and it achieves an equilibrium position during the `wetting/equilibrium phase.' Sometimes, a Rayleigh jet (shown in Fig. \ref{fig:droplet_images}) is observed due to excess kinetic energy at the end of retraction. More details on the `relaxation' and `wetting/equilibrium phase' can be found in literature \citep{yamamoto2016droplet,richard2002contact,antonini2013drop}. 

\subsection{Impact on oscillating substrates}
\label{result_osc}

\begin{figure}
\centering
    \includegraphics[width=1\textwidth]{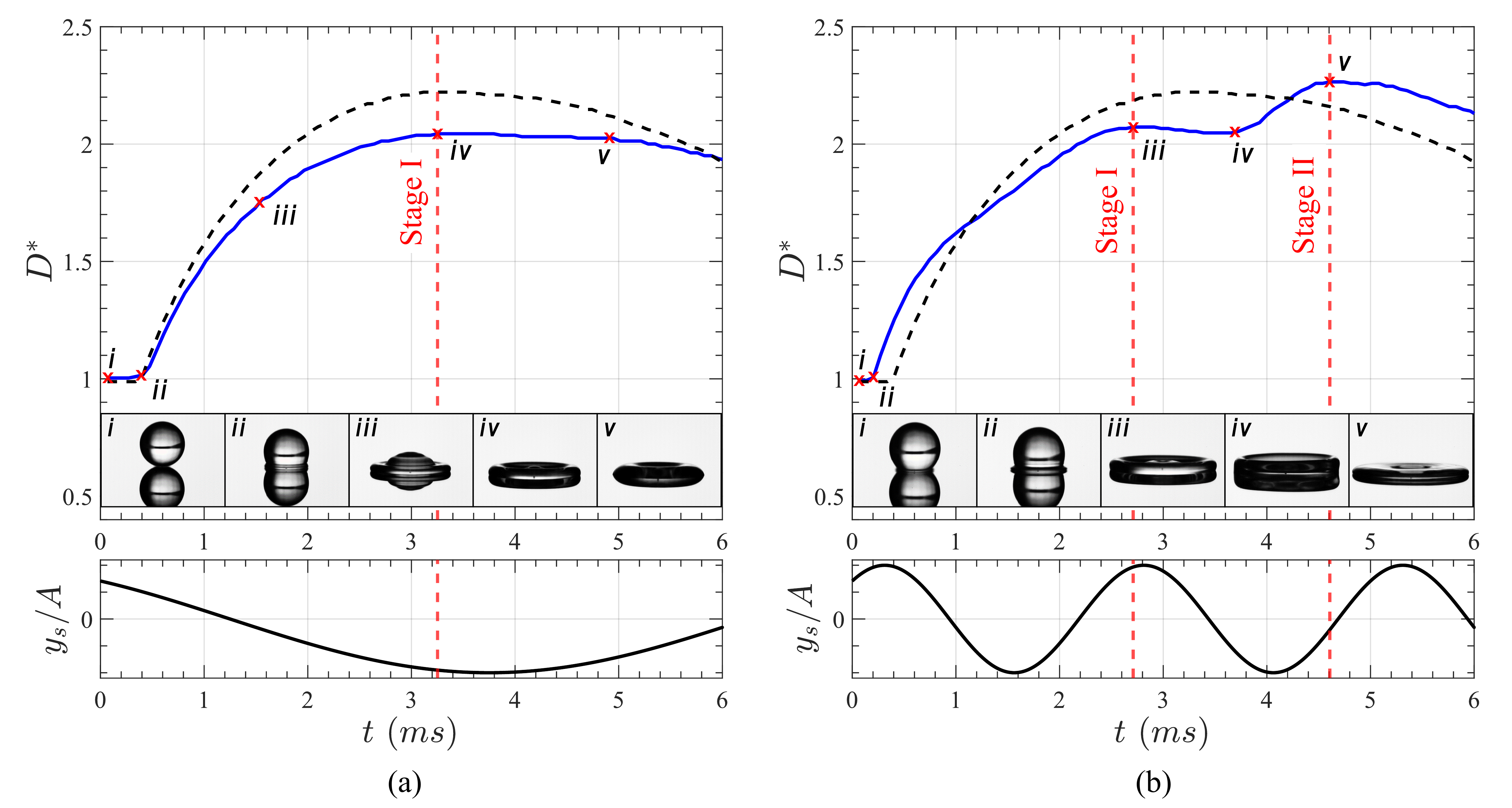}
\caption{ Temporal evolution of post-impact normalized droplet diameter ( $\text{D}/\text{D}_0$) with snapshots illustrating droplet profiles (Fig. \ref{fig:droplet_images}) for $We = 27$, $Re = 2300$ and (a) $\text{f}$ = $100\text{Hz}$, $\text{A}$ = $0.25mm$ and $\phi = 3 \pi/4~rad$ (b) $\text{f}$ = $400\text{Hz}$, $\text{A}$ = $0.125mm$ and $\phi = \pi/4~rad$; The bottom plots show the evolution of substrate motion ($y_{s}/A = sin(2 \pi f t + \phi)$) for both cases. -- {--} {--} shows the spreading for impact on a static surface. }
\label{fig:spreadprofile}
\end{figure}

\begin{figure}
    \centering
    \includegraphics[width=1\textwidth]{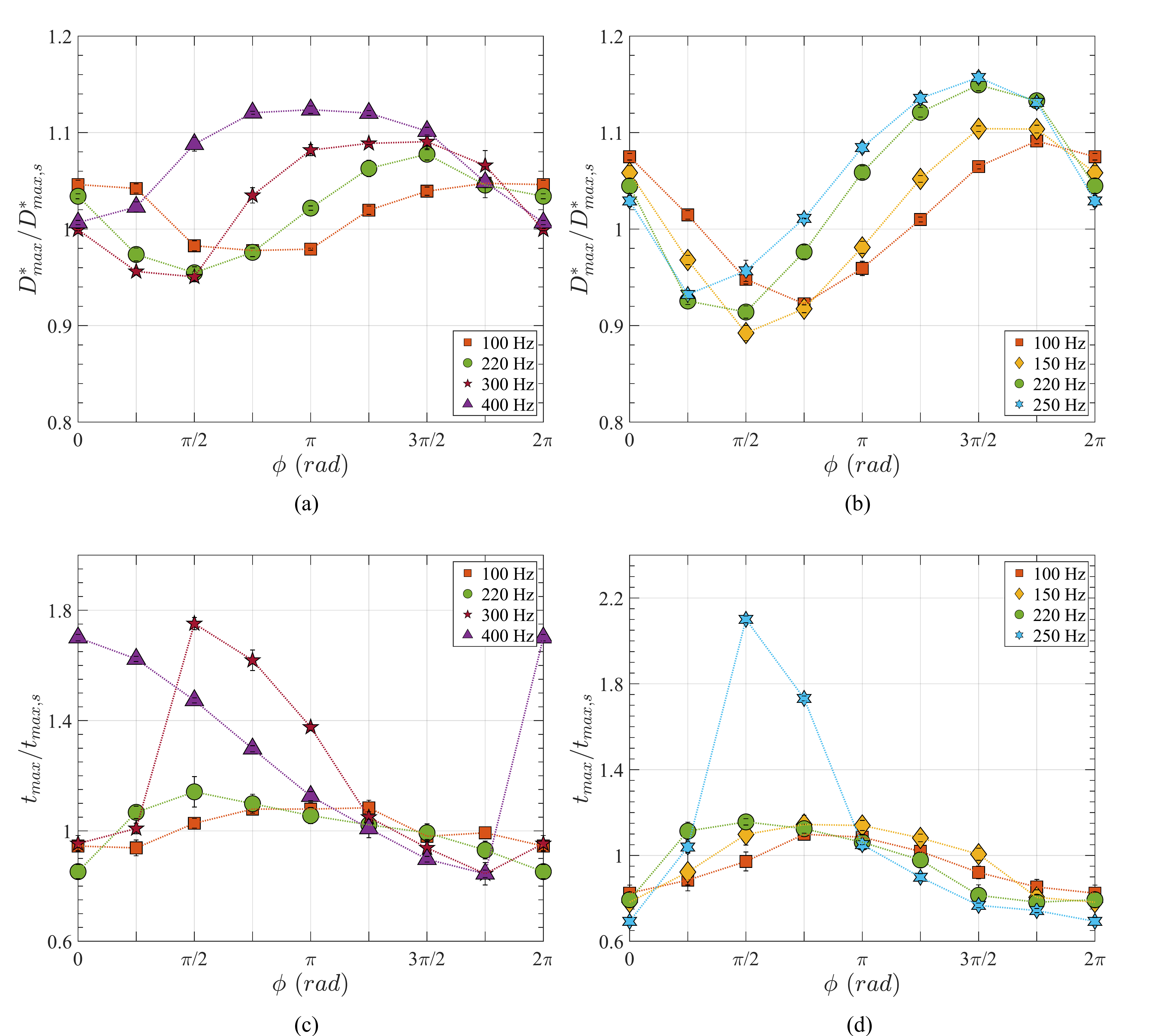}
\caption{(a,b) Normalized maximum spread, $\textit{D}^*_{max}/\textit{D}^*_{max,s}$ as a function of phase, $\phi$ at impact for various frequencies for $We \approx 27$, and $Re \approx 2300$. (a) $A = 0.125~mm$ (b) $A = 0.25~mm$. Here the error bars represent the extent of standard deviation about the mean value. (c,d) Normalized time to maximum spread, $\textit{t}_{max}/\textit{t}_{max,s}$ as a function of phase, $\phi$ at impact for various frequencies. (c) $A = 0.125~mm$ (d) $A = 0.25~mm$.  The significantly higher values of $\textit{t}_{max}/\textit{t}_{max,s}$ seen in both figures are a consequence of \textit{Stage-II} spreading. Here the error bars represent the extent of standard deviation about the mean value.}
\label{fig:Dmax_phase_freq}
\end{figure}

We will now present the experimental results for droplet impact on oscillating substrates. For these experiments, the substrate was actuated with a sinusoidal wave, and the phase at the impact ($\phi$) was controlled using the laser-triggered delay generator as detailed in Sec. \ref{sec:expt_setup}. The post-impact stages of droplet spread on an oscillating substrate qualitatively resemble that on a static one, as shown in Fig. \ref{fig:droplet_images}b and c. However, the spreading time and the maximum spread change due to the continuous movement of the substrate. The dynamics of post-impact spreading for impact on oscillating substrate, thus, depends on the motions of both the droplet and the substrate. Their combined effect is quantified by the relative droplet velocity, defined as $V_{rel} = V_{drop} + V_{s}$, which changes with time. Here, $V_{drop}$ is the instantaneous downward velocity of the droplet. It is to be noted that based on the phase at impact ($\phi$), the substrate velocity ($V_s$) at impact can be upward ($0<\phi < \pi/2~ \& ~3\pi/2<\phi<2\pi$) or downward ($\pi/2 < \phi < 3\pi/2$), thereby increasing or decreasing the relative impact velocity ($V_{rel}$), respectively. Let's elaborate on this using the case where the substrate oscillated with amplitude, $A = 0.25~mm$, frequency, $f = 100\text{Hz}$, and phase at impact was $\phi = 3 \pi/4$. 
The temporal evolution for this impact is shown in Fig. \ref{fig:spreadprofile}a. The top panel of the figure illustrates the instantaneous droplet diameter during the spreading processes. The solid line represents the impact on the oscillating substrate, and the dotted line refers to the impact on the static substrate at the same $We$. The morphology of the deformed droplets is shown for some key instances as insets. The lower panel displays the instantaneous locations of the substrate during the spreading process. 
Figure \ref{fig:spreadprofile}a shows that, at the time of impact, the substrate moves downwards, away from the droplet with velocity $\text{V}_{s} \approx -0.11~m/s$. 
Furthermore, after the impact, $\text{V}_{s}$ remains negative as the substrate continues with a downward motion. Naturally, the droplet experiences a lower value of $V_{rel}$ throughout the duration of spreading. The decreased $V_{rel}$ induces reduced vertical compression, and hence a reduced spreading ($D_{max}^* = D_{max}/D_0=2.04$), compared to the spreading for impact on the static substrate ($D_{max,s}^* = 2.22$) for the same impact velocity or $We$ as seen in Fig. \ref{fig:spreadprofile}a. The change in $V_{rel}$ also affects the time taken by the droplet to reach maximum spreading ($t_{max} = 3.3~ms$ and $t_{max,s} = 3.0~ms$ for impacts on oscillating and static substrates, respectively).
The observed dynamics quantitatively change with $\phi$ because the instantaneous $V_s$ and hence, $V_{rel}$ follow different temporal histories leading to either inhibited or assisted spreading of the droplet. The modification in spreading diameter and time for spreading, due to substrate oscillation is illustrated in Fig. \ref{fig:Dmax_phase_freq}, where we compared $D^*_{max}/D^*_{max,s}$ and $t^*_{max}/t^*_{max,s}$  vs. $\phi$ for different $A$ and $f$ for a fixed $We (\approx 27$) and $Re(\approx 2300$).

From Fig. \ref{fig:Dmax_phase_freq}a and b it is evident that the oscillations rendered spreading difficult compared to the impact on the static substrate ($D_{max}^*/D_{max,s}^*\ < 1$), for $0 \lessapprox \phi \lessapprox \pi$. On the other hand, the oscillations enhanced the spreading ($D_{max}^*/D_{max,s}^* > 1$) for $\pi \lessapprox \phi \lessapprox 2\pi$. 
At lower amplitude ($A=0.125~mm$), for the lowest frequency ($f=100\text{Hz}$), the minimum $D_{max}^*$ occurs close to $\phi = \pi$ and for higher $f$s the minimum $D_{max}^*$ condition shifts towards lower $\phi$ values ($\pi /2$ for $f=220\text{Hz}$ and $f=300\text{Hz}$) as shown in Fig. \ref{fig:Dmax_phase_freq}a. On the other hand, the maximum $D_{max}^*$ occurs close to $\phi = 2\pi$ for $f=100\text{Hz}$ and with increase in $f$, it shifts to lower $\phi$ ($3/2 \pi$ for $220\text{Hz}$). Fig. \ref{fig:Dmax_phase_freq}a also shows that the value of the maximum $D_{max}^*$ across various $\phi$ monotonically increases with $f$. However, the value of the minimum $D_{max}^*$ first decreases with an increase in frequency until $f\lessapprox 150~\text{Hz}$, then increases for higher frequencies. These trends are qualitatively in close agreement with previous numerical studies by \citet{moradi2020numerical,lin2022dynamic,li2022hydrodynamic}.
However, we also observe a modified behavior for higher frequencies ($f\geq 250\text{Hz}$) in that the enhancement in spreading ($D_{max}^*/D_{max,s}^*>1$ ) sustained for an increasingly large range of $\phi$ values. For example, we observed $D_{max}^*/D_{max,s}^* > 1$ across all $\phi$s at $A = 0.125~mm$ and $f= 300~\text{Hz}$ (Fig. \ref{fig:Dmax_phase_freq}a). 

At low frequencies, in the duration of droplet spreading, the substrate completes only a partial cycle of oscillation, and depending on the direction of the motion, the spreading is assisted or inhibited. We generally observe a single local maximum in the instantaneous droplet diameter. This is identified as \textit{Stage I} spreading in Fig. \ref{fig:spreadprofile}a, which shows the dynamics for $100\text{Hz}$.
On the other hand, at higher frequencies, the substrate undergoes multiple cycles of oscillations during the spreading process (Fig. \ref{fig:spreadprofile}). 
Here, the initial spread is governed by the first cycle of oscillation where the $D^*$ profile reaches the first maxima, $\textit{Stage-I}$. The droplet, subsequently, initiates a 'relaxation phase' as $D^*$ starts reducing. But, before the retraction phase can occur, the substrate starts a downward acceleration, causing the rim of the droplet to increase in height as the substrate rapidly moves downwards. This is evident in snapshots for $400\text{Hz}$ shown in Figs. \ref{fig:droplet_images}b and \ref{fig:spreadprofile}b. The heightened rim subsequently collapses as the substrate starts an upward movement during the next cycle of oscillation (Fig. \ref{fig:spreadprofile}b). This yields an increase in instantaneous $D^*$, creating a second local maximum. This is termed as \textit{Stage-II} spreading as shown in Fig. \ref{fig:spreadprofile}b. For some $\phi$s, the overall maximum spreading is observed during the \textit{Stage-II}, as shown in Fig. \ref{fig:spreadprofile}b. 
Since the maximum spread is achieved at a later cycle of oscillation, the  $\textit{t}_{max}/\textit{t}_{max,s}$ for the \textit{Stage-II} maximum spread is significantly higher, as illustrated in Figure. \ref{fig:Dmax_phase_freq}c and d.  

The spreading dynamics ($D_{max}^*/D_{max,s}$) remain qualitatively similar at various phases across the different amplitudes. More specifically, we observe the maximum and minimum spread ($D_{max}^*$) to occur at the same $\phi$ for different amplitudes (Fig. \ref{fig:Dmax_phase_freq}a vs. b). However, the minimum and maximum values of $D_{max}^*$ are reduced and amplified with an increase and decrease in $A$, respectively. Clearly, variations in amplitude do not alter the time scales associated with substrate oscillation but affect the magnitude of the effect or $D_{max}^*/D_{max,s}^*$. Comparison of $D_{max}^*/D_{max,s}^*$ for a larger range of $A$ are shown in the supplementary materials. We have also performed experiments with three different $We$ (and $Re$), which results in qualitatively similar behavior. The $D^*_{max}$ values from these experiments can be found in the supplementary materials.


\section{Scaling analysis}
\subsection{\textit{Stage-I} spreading}
\subsubsection{Time for maximum spread}
\label{sec:t_max}

As the \textit{Stage-I} spreading is kinematically controlled for an inertia-driven impact, the characteristic timescale governing the spreading dynamics of a droplet impacting on a substrate is the crashing time ($\tau$). Physically, it is the time taken by the tip of the droplet to reach the substrate in the absence of any form of deceleration, and its relation with droplet diameter can be expressed as 
\begin{equation}
D_0=\int_{0}^{\tau} V_{rel}(t)dt.
\label{eq:tau_def}
\end{equation}
Imposing $V_{rel}=V_0$, for impact on static substrates (i.e., no substrate motion), the crashing time can be shown to be $\tau_s = \text{D}_0/\text{V}_0$.
For a wide range of impact conditions ($2<We<900$), \citet{clanet2004maximal} proved that $\tau_s$ is indeed the relevant timescale. 
The time for the droplet to achieve maximum spreading ($t_{max}$), however, is different than the crashing time ($\tau$). For impact on static substrates, it has been shown that, although the droplet achieves a significant part of its total deformation and spreading at $t\le\tau_s$, the droplet still possesses a small amount of kinetic energy which decays almost asymptotically \citep{roisman2002normal}. This remnant energy causes further deformation and spreading, albeit at a much weaker rate compared to $t\le\tau_s$. Viscous loss and capillarity dominate at this stage ($t>\tau_s$), eventually restricting and stopping the spreading process. One can impose an assumption of $t_{max,s}=\tau_s$, as it was done for several studies. However, this assumption will lead to an under-prediction of $t_{max}$ compared to experimental measurements. This introduces significant errors in evaluating viscous losses while the modeling of droplet spread. The accurate estimation of $t_{max,s}$ from simple scaling is challenging, which is also recognized by other studies \citep{du2021analytical}. To get a realistic estimate of $t_{max,s}$, we plotted the ratio of ${{t}_{max,s}}/{\tau_s}$ for impact on solid substrates as function of $We$ in Fig. \ref{fig:0HzWe_tmaxfit}b and evaluated a linear relation with $We$, ${{t}_{max,s}}/{\tau_s} =\ 0.0109We + 1.0346$. As the overall $We$ does not change due to the substrate oscillation, we will assume that this relationship between crashing time, time for the maximum spread, and $We$ holds for impact on oscillating substrates and hence,
\begin{equation}
\frac{{t}_{max}}{\tau} =\ 0.0109\ We\ +\ 1.0346.
\label{eq:t_max_fit_osc}
\end{equation}
The crashing time for impact on oscillating substrates ($\tau$) is affected by the motion of the substrate and hence, is different from its counterpart of impact on static substrates ($\tau_s$). For impact on oscillating substrates, $\tau$ is a function of impact velocity ($V_0$), and frequency ($f$), amplitude ($A$), phase ($\phi$) of the oscillation, and can be evaluated by using $V_{rel}=V_0 + 2 A\pi f\cos{(2 \pi f t + \phi)}$ in Eq. \ref{eq:tau_def}. Here, it is assumed that the downward velocity ($V_0$) of the droplet with respect to a lab-fixed reference is constant for the crashing period, i.e., $0<t\leq \tau$. In supplementary material, we compared the trajectory of the tip of the impacting droplet and the instantaneous droplet spread, which shows that the droplet descends with nearly a constant velocity until $t=\tau$, justifying the assumption. Once the crashing time ($\tau$) is calculated theoretically (Eq. \ref{eq:tau_def}), we use the correlation in Eq. \ref{eq:t_max_fit_osc} to evaluate the theoretical time taken for maximum spread during droplet impact on oscillating substrates. 

\begin{figure}
    \centering
    \includegraphics[width=1\textwidth]{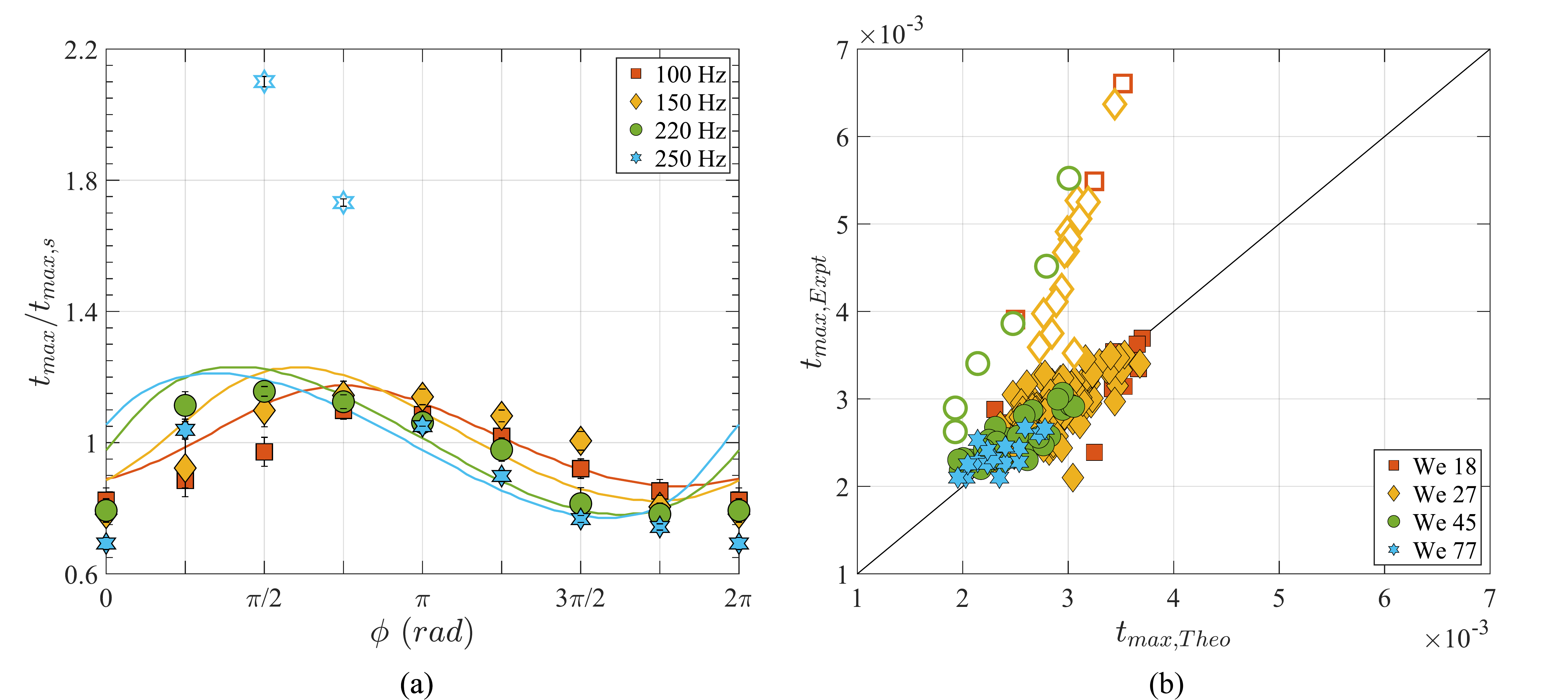}
\caption{(a) Comparison of experimental and theoretically predicted values for normalized maximum spreading time, $\textit{t}_{max} / \textit{t}_{max,s}$, as a function of phase at impact, $\phi$, for $We\ =\ 27$, and $.25~mm$ for various frequencies. (b)  Comparison of experimental and theoretically predicted the maximum spreading time, $\textit{t}_{max}$. The filled symbols represent $\textit{Stage-I}$ spreading and open symbols are for $\textit{Stage-II}$ as seen from experiments.  The plots show that theoretical values for $\textit{t}_{max}$ show a good match for all data showing $\textit{Stage-I}$ spreading.}
\label{fig:t_max_compare_theo}
\end{figure}

Figure \ref{fig:t_max_compare_theo}a shows the theoretical (solid lines) and experimental (symbols) normalized $t_{max}$ for impacts on oscillating substrates for various $\phi$ and $f$ at $We = 27$, $Re = 2300$, $A = 0.25~mm$. For most conditions, the theoretical values show good agreement with the experimental data in that it captures both qualitative and quantitative changes with $\phi$ and $f$. Large discrepancies were observed for the conditions (e.g., $f=250\text{Hz}$) for which the maximum spreading was obtained during the subsequent oscillations of the substrate (\textit{Stage-II} spreading). The analyses of crashing time that led to Eq. \ref{eq:tau_def} account for the inertia controlled \textit{Stage-I} spreading, but discount the additional spreading which occurs during the retraction stage of the droplet by the action of subsequent oscillations (\textit{Stage-II}). 

In Fig. \ref{fig:t_max_compare_theo}b, we compare the theoretical vs. experimental $t_{max}$ for all conditions studied in our experiments covering a $We$, $A$, $f$, and $\phi$. Again, we observe good agreement between the experiment and theory for all the conditions with \textit{Stage-I} spreading with an error range of $\pm10\%$. As expected, the conditions affected by the \textit{Stage-II} spreading (marked with open symbols) show discrepancies due to the above-mentioned reason. 
We will derive scaling for \textit{Stage-II} spreading later in Section \ref{sec:sec_stage}. 

\begin{figure}
    \centering
    \includegraphics[width=0.7\textwidth]{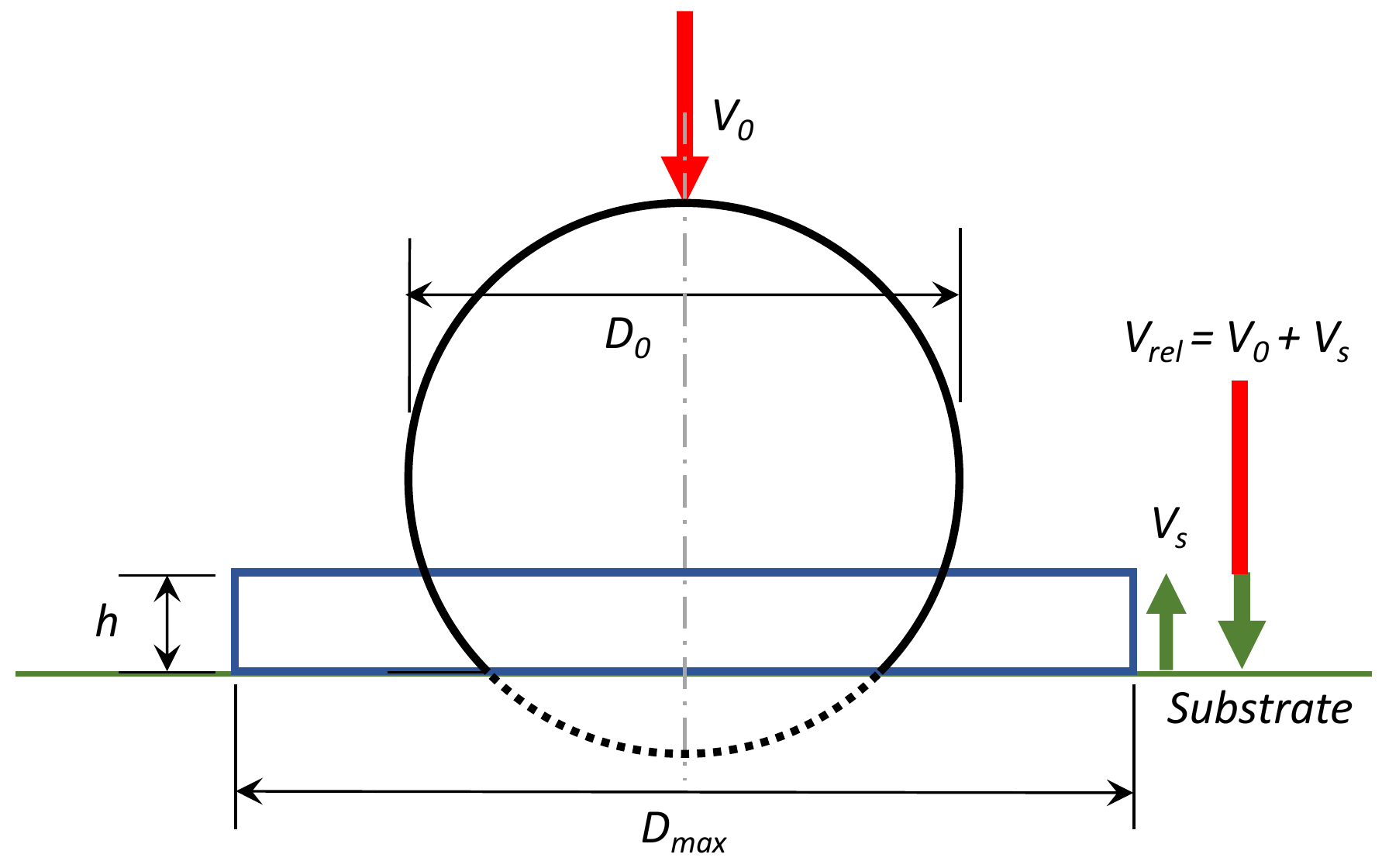}
\caption{Illustration of mass-conservation during droplet impact; $\text{D}_0$ \& $\text{V}_0$ are initial droplet diameter and velocity, $\text{D}_{max}$ \& $h$ are dimensions of spreading droplet, $\text{V}_{s}$ is the velocity of substrate, and $\text{V}_{rel}$ is the relative velocity.}
\label{fig:mass_consv}
\end{figure}

\subsubsection{Maximum spreading diameter}
\label{sec:energy_balance}
Recognizing that we have an good prediction of $t_{max}$ for impact on oscillating substrates, we next proceed to analyze the energy balance to obtain a theoretical expression of normalized maximum spreading diameter ($D^*_{max}$).
As mentioned before, for impacts with relatively high $We$ and ${Re}$ the spreading is inertia driven. The hydrophobic nature of the substrate ($\theta_{eq} = 90^{\circ}$) allows the droplet to spread with the effects of viscosity being confined mostly to the boundary layer close to the substrate. Thus, the dynamics of spreading is an outcome of the balance between the initial kinetic energy ($\text{KE}$), which in parts, is converted to the surface energy ($\text{SE}$) of the droplet, and is lost to viscous dissipation ($\text{W}$).
It is generally assumed that the droplet has negligible kinetic energy at maximum spreading. Furthermore, for impacts on oscillating substrates, the total energy in the system is altered by the presence of an oscillating boundary, which injects and withdraws energy from the system as the substrate moves upwards and downwards, respectively. The energy balance between the states of the droplet before impact and at the maximum spread, then, can be expressed as
\begin{equation}
    KE_{0} + SE_{0} + E_{s} =\ SE_{f} + W.
\label{eq:energy_bal_osc}
\end{equation}
Here, the kinetic and surface energies of the droplet before the impact are $\text{KE}_{0}= (1/12)\pi\rho D^3_0V_0^2$ and $\text{SE}_{0}=\gamma \pi D_0^2$, respectively.
To simplify the analysis, it was assumed that the geometry of the deformed droplet at the end of the spreading resembles that of a shallow cylinder (or pancake) of diameter, $D_{max}$ and `splat height' $h$ (shown in Fig. \ref{fig:mass_consv}). Although the droplet takes a far more complicated shape with rims during the impact, \citet{eggers2010drop} showed that rim formation occurs during the retraction phase. This justifies the assumption of a cylindrical geometry at the instant of maximum spread before the retraction begins.
With this assumption, the final surface energy can be expressed as $\text{SE}_{f} = \gamma \pi\left[D_{max}^2/4 + D_{max} h\right]$. From mass conservation, the height of the deformed droplet can be shown to be $h = (2/3)\ (D_0^3 / D^2_{max})$. 
The overall viscous dissipation is estimated by $W = \int^{t_{max}}_{0} \int^{}_{\Omega} \Phi d\Omega dt \approx \Phi \Omega t_{max}$.
Here, $\Phi$ is the viscous dissipation rate per unit volume and can be approximated as $\Phi \sim \mu \left({\text{V}_0}/{\delta}\right)^2$.
The boundary layer thickness is obtained by the relation $\delta\sim {D_0}/\sqrt{Re}$ \citep{white2006viscous}.
The volume of the boundary layer at the bottom of the cylindrical droplet, where the viscous loss is significant, is given by $\Omega={\pi D_{max}^2 \delta}/{4}$. 
Since the expressions for evaluating $\Phi$ and $\delta$ are approximations, we introduce $\alpha$ as the scaling factor to accurately assess the total viscous loss in the boundary layer, 
\begin{equation}
  W = \alpha \frac{\rho\, V_0^3}{\sqrt{Re}} \pi D_{max}^2 t_{max}.
\label{eq:W_simplified}
\end{equation}
The complexity of quantitative assessment of viscous dissipation in droplets during impact on substrates is well-known. A scaling factor to account for the discrepancies has been proposed previously by \citet{eggers2010drop}. The process of evaluation for $\alpha$, in our study will be discussed later.

To account for the additional energy supplied by the moving substrate, we introduce $E_s$, defined as the total energy imparted by the surface to the droplet until maximum spread is achieved ($t=t_{max}$). 
Assuming the whole droplet is moving with respect to the substrate, we can express $E_{s}=(\rho \pi/8) \int^{t_{max}}_{0} (d\Vol/dt) V_{s}^2(t') dt'$. 
Here, we assume that the contact length of the droplet with the surface scales with the droplet length-scale, i.e., the initial diameter $D_{0}$. Thus, the rate of change of the volume of liquid affected by the surface movement can be expressed as $(d\Vol/dt)\approx D_0^2 V_s$. These leads to
\begin{equation}
    E_{s}\ \approx \rho\, \frac{\pi\, D_0^2}{8}\, \left(\,\int^{t_{max}}_{0}   V_{s}^3(t')\, dt'\, \right).
\end{equation}
Here the energy transfer from the substrate to the droplet is assumed to be lossless. 
We note that the above expression for $E_s$ accounts for the `directionality' of the energy transfer. $E_s$ is positive and negative when energy is added to and withdrawn from the droplet, respectively. 
By substituting the expressions for various forms of energies in Eq. \ref{eq:energy_bal_osc} and by normalizing the lengths by $D_0$, and time by $\tau$, we find
\begin{dmath}
\frac{We}{12}\ +\ 1\ +\ \frac{E_s}{\gamma\ \pi\ D_0^2}\ =\ \frac{(D^*_{max})^2}{4}\ +\ D^*_{max}\, h^*_{max}\ +\ \alpha\ \frac{We}{\sqrt{Re}}\ (D^*_{max})^2\ \frac{t_{max}}{\tau}.
\label{eq:D_max_final}
\end{dmath}

\begin{figure}
    \centering
    \includegraphics[width=0.5\textwidth]{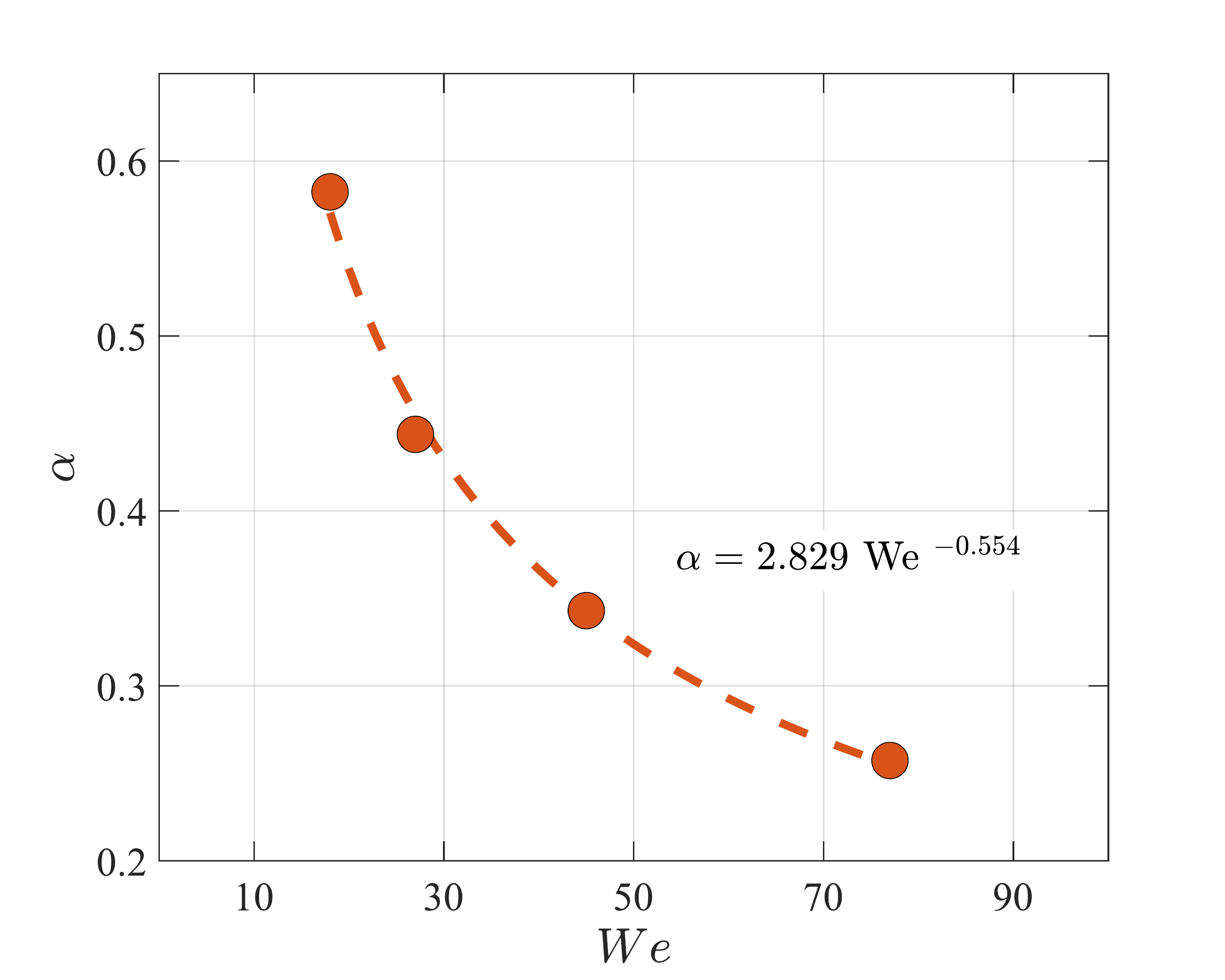}
\caption{Empirical fit for $\alpha$ expressed in \ref{eq:D_max_final} obtained from experimental values for static impact for $We\ =\ 18,\ 27,\ 45\ \&\ 77$}
\label{fig:alpha_fit}
\end{figure}

Theoretical estimation of $\alpha$, the scaling factor arising from the boundary layer analyses, is highly sensitive to the assumptions and simplifications. Since an accurate calculation of the scaling factor is difficult from theory, we use experimental data. Here, we utilize the data from the experiments with impacts on static substrates ($E_s=0$). By substituting the measured $D^*_{max}$ and $\tau$ in Eq. \ref{eq:D_max_final} for various impact conditions, we can solve for $\alpha$. As shown in Fig. \ref{fig:alpha_fit}, the best-fit $\alpha$ depends on the impact conditions as $\alpha = 2.829 We^{-0.55}$. A similar approach was also successfully used by \citet{eggers2010drop} for their analyses of impacts on static substrates.

\begin{figure}
    \centering
    \includegraphics[width=1\textwidth]{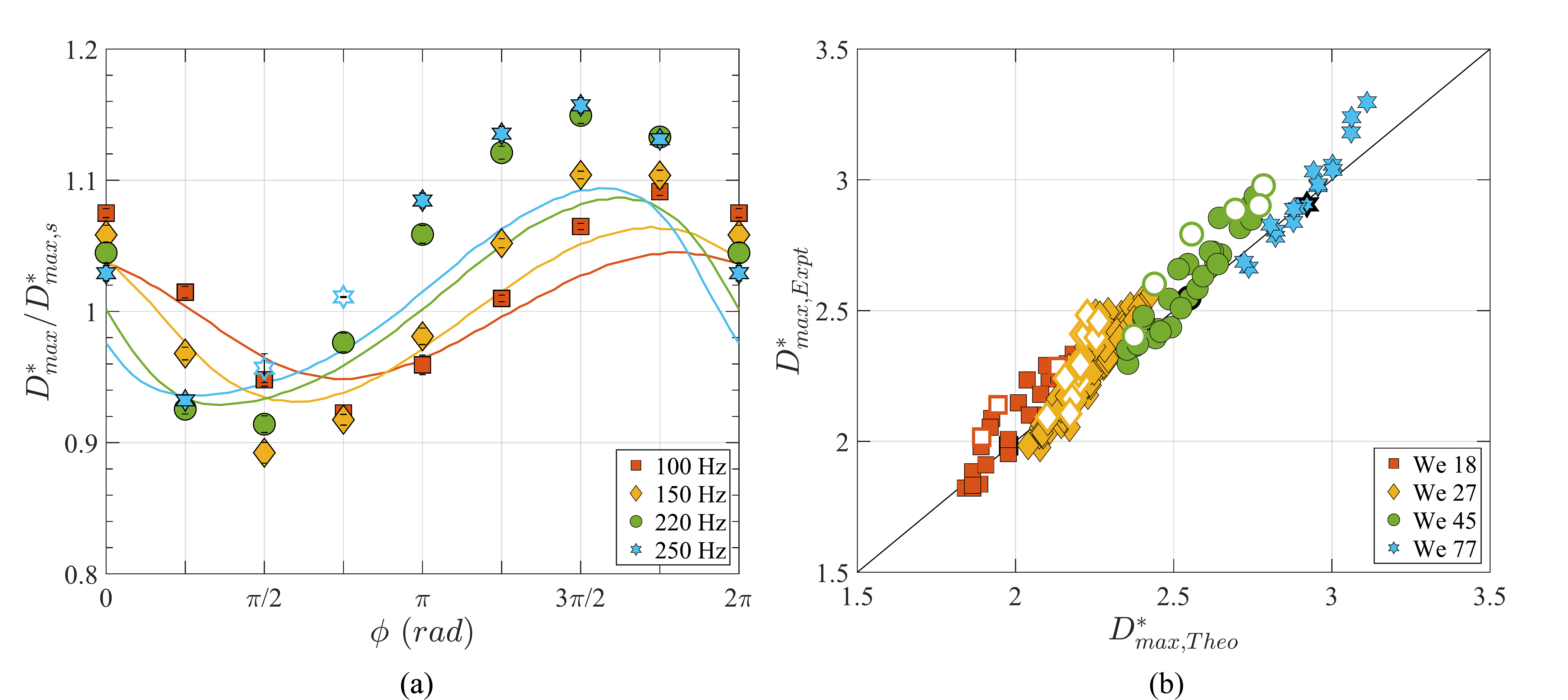}
\caption{(a) Comparison of experimental and theoretically predicted values for $D^*_{max}/D^*_{max,s}$ for $We = 27$, $Re = 2300$, $A = 0.25~mm$ as a function of $\phi$. (b) Comparison of experimental and theoretically predicted absolute values for $D^*_{max}$ at the instant of maximum spreading for all experimental data. The filled icons represent $Stage~1$ spreading, and unfilled icons are for \textit{Stage II} as seen from experiments. }
\label{fig:Dmax_compare_theo}
\end{figure}

Equation \ref{eq:D_max_final} can be solved to evaluate $D^*_{max}$ for impacts on oscillating substrates with various amplitudes, frequencies, and phases. Figure \ref{fig:Dmax_compare_theo}a compares the experimental (symbols) and theoretical (lines) $D^*_{max}$ for $We = 27$, $Re = 2300$, and a fixed value of amplitude ($A=0.25mm$) over a range of frequency ($f$) and phase ($\phi$). We notice that the model predicts the decrease and increase in $D^*_{max}$ as frequency and phase change with reasonable accuracy. Furthermore, Fig. \ref{fig:Dmax_compare_theo}b compares $D^*_{max}$ measured in all the experiments with the corresponding theoretical predictions. Overall, good qualitative and quantitative agreements have been observed across all the conditions. 

We can make a few additional observations from the analyses and the results of $D^*_{max}$.
By using representative values, one can show that $E_s$ is relatively insignificant compared to the other terms ($KE_{0}$, $SE_{0}$, $SE_{f}$ and $W$) in energy balance  equation (Eq. \ref{eq:energy_bal_osc}). This implies that the modification in a maximum spread of the droplet for impact on the oscillating substrate does not directly come from the additional energy introduced by the substrate. The oscillations also modify the droplet's relative velocity, affecting the history of the deformation process and, thus, spreading time, $t_{max}$. This, in turn, affects the viscous dissipation, $W$, and hence $D_{max}$. Thus, it is essential that $t_{max}$ is modeled accurately. Furthermore, an order of magnitude analysis of the other terms in Eq. \ref{eq:D_max_final} shows that about $40-60\%$ of total initial energy ($KE_{0}+SE_{0}$) is lost through viscous dissipation ($W$). This large viscous dissipation was also reported for droplet impact on static substrates by \citet{wildeman2016spreading} and \citet{li2022hydrodynamic}. 

\begin{figure}
    \centering
    \includegraphics[width=0.95\textwidth]{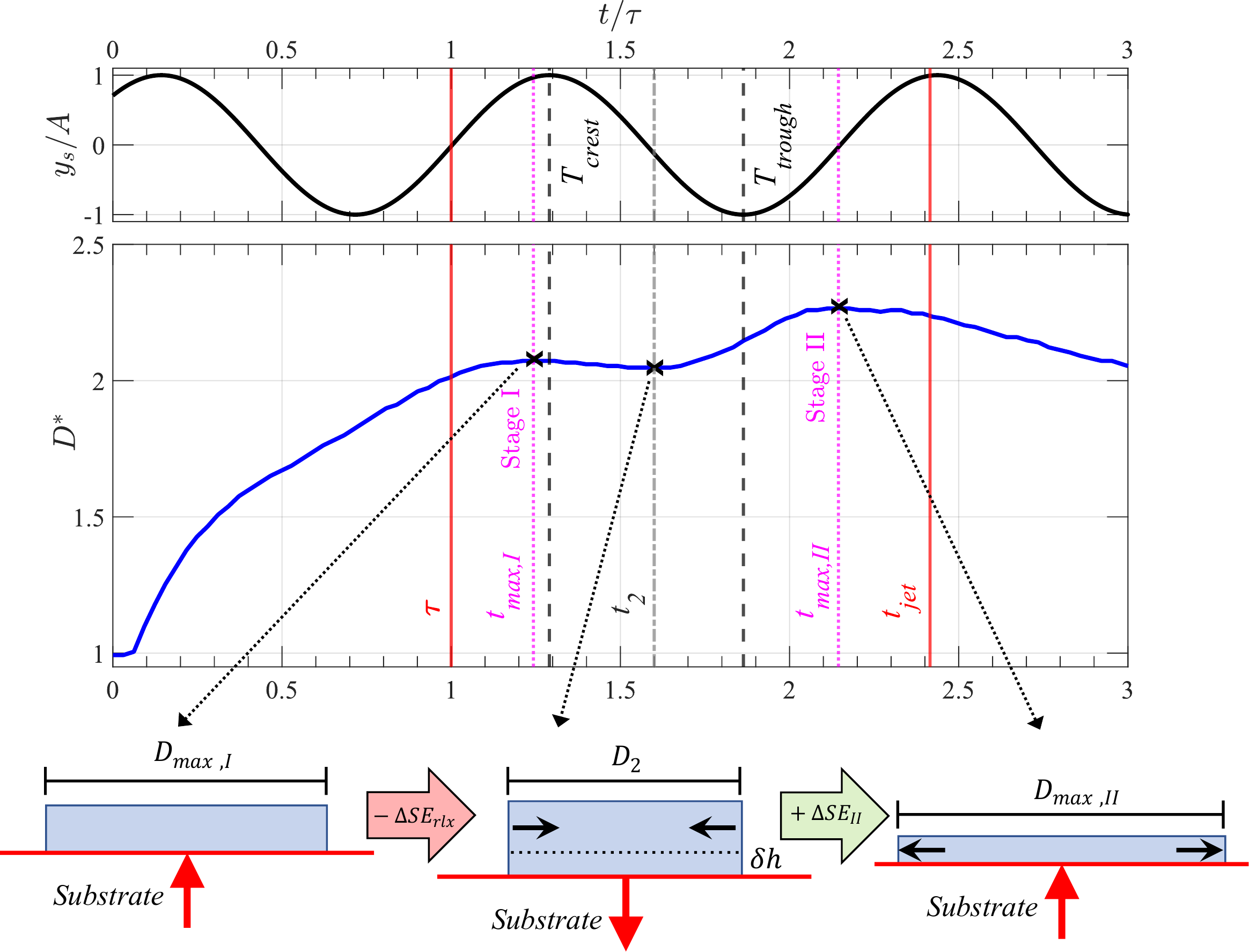}
\caption{Temporal evolution of post-impact normalized droplet diameter ($\text{D}/\text{D}_0$) illustrating the important time instants for $\textit{Stage-I}$ and $\textit{Stage-II}$ spreading for $We = 27$, $Re = 2300$, $\text{f}$ = $400\text{Hz}$, $\text{A}$ = $0.125mm$ and $\phi = \pi/4~rad$. The top plot shows the evolution of substrate motion ($y_{s}/A = sin(2 \pi f t + \phi)$). The illustration showcases the mechanism of \textit{Stage II} spreading.}
\label{fig:sec_stage_illustration}
\end{figure}

\subsection{\textit{Stage-II} spreading}
\label{sec:sec_stage}
As discussed earlier and seen in Figures \ref{fig:spreadprofile}b, and  \ref{fig:Dmax_phase_freq}, an impacting droplet on an oscillating substrate may attain its maximum spread during the retraction process due to the subsequent cycle of oscillations. If the condition is conducive, the oscillation of the substrate may assist the droplet in attaining a local instantaneous diameter greater than \textit{Stage-I} spreading, which occurs immediately after the impact (discussed and analyzed in the previous section). We will now discuss the physical mechanism for \textit{Stage-II} spreading and the conditions required for $D_{max}$ to occur during this period.

Physically, at the point of maximum spread (\textit{Stage I}), the droplet loses its kinetic energy and contains excess surface energy in the form of a deformed shape. In the absence of 
substrate motion, the droplet will undergo a relaxation phase, where the diameter will reduce, and the height will increase. It will eventually form a jet-like central liquid column (Fig. \ref{fig:droplet_images}a). The time for this jet formation ($t_{jet}$) is primarily controlled by the capillary process and, in our experiments (shown in supplementary material), is found to be 
\begin{equation}
    t_{jet} \approx \frac{1}{2}\ t_{cap}, 
\label{eq:t_jet}
\end{equation}
where $t_{cap} = \sqrt{{\rho D_0^3}/{\gamma}}$ is the capillary time. \citet{yamamoto2016droplet} also reported similar findings for their study with droplet impact on the stationary superhydrophobic substrate. The relaxation of the droplet can be approximated as a capillary process, and the instantaneous diameter during the relaxation ($D_2$) can be expressed as 
\begin{equation}
    D_2\approx D_{max,I} \cos(2 \pi \Delta t / t_{cap}'),
    \label{eq:D_relax}
\end{equation}
where the $\Delta t$ is the time elapsed after maximum spread occurred at $t=t_{max,I}$, and $D_{max,I}$ is the diameter at the maximum spread at \textit{Stage I}. 
Assuming that the droplet keeps its cylindrical shape during the relaxation period, the reduction in the surface area ($\Delta SE_{rlx}= SE_{f}-SE_{rlx}$) 
can be expressed as
\begin{equation}
\Delta SE_{rlx}= \gamma \pi \left( \frac14 (D_{max,I}^2-D_{2}^2) + \frac23 D_0^3\left(\frac{1}{D_{max,I}}-\frac{1}{D_2}\right) \right).
\label{eq:SE_relax}
\end{equation}

\begin{figure}
    \centering
    \includegraphics[width=0.5\textwidth]{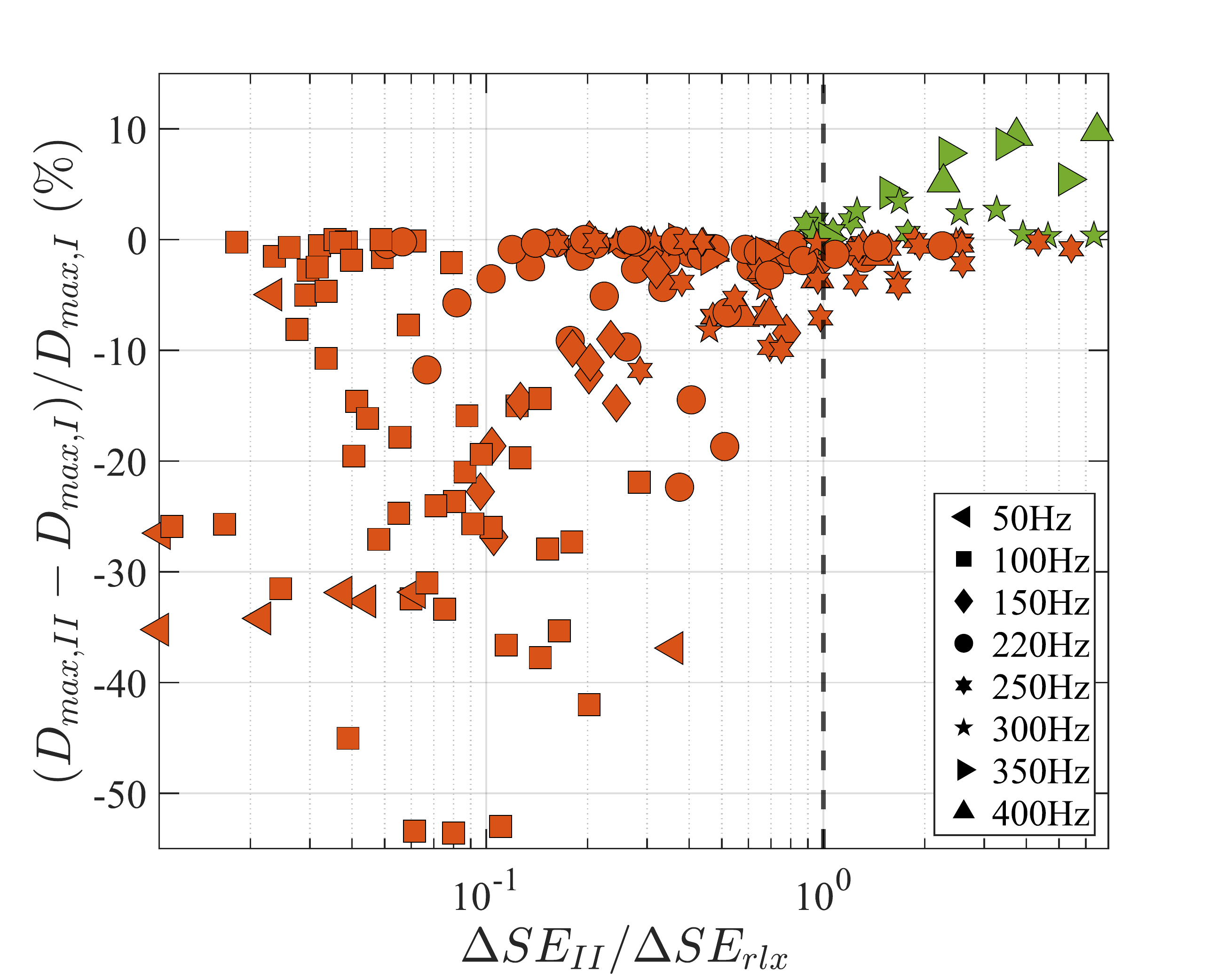}
\caption{Illustration of energy criteria for $\textit{Stage-II}$ spread. Experimentally observed percentage change in droplet spread diameter from $\textit{Stage-I}$ and $\textit{Stage-II}$ as a function of $\Delta SE_{II}/\Delta SE_{rlx}$ for all cases. The dotted line $\Delta SE_{II}/\Delta SE_{rlx} = 1$ marks the threshold for the occurrence of maximum spread at $\textit{Stage-II}$, differentiated by the green colored symbols.}
\label{fig:D2minusD1vsSE}
\end{figure}

To achieve the maximum spread during \textit{Stage-II}, the substrate oscillation must provide additional energy to overcome the reduction in surface area during the relaxation process, thereby causing a greater spread. From observation, it was noted that this is achieved when the phase of the substrate oscillation at the droplet impact is such that the substrate initiates the downward motion of its sinusoidal trajectory during the relaxation phase of the droplet. 
This sudden downward movement of the substrate momentarily pulls the adjacent (bottom) part of the droplet while the top part undergoes the relaxation process. This process lasts until the top of the droplet feels the downward motion of the substrate, and the whole droplet starts moving downwards. Although short, this process causes a momentary increase in droplet height ($\delta h$) during the relaxation process. Recognizing that the disturbance from the bottom to the top of the droplet is transported through capillary waves along the droplet surface, the increase in droplet height can be estimated as $\delta h\approx V_s t'_{cap}$. Here, $t'_{cap}=\sqrt{\rho D_{max,I}^3/\gamma}$ the capillary time for the deformed droplet at the end of \textit{Stage-I} spreading.
Once the substrate reaches its lowest position and subsequently reverses the direction of its motion ($t=T_{trough}$), the elongated droplet collapses, thereby causing the droplet to undergo \textit{Stage-II} spreading. The process, with the help of instantaneous droplet diameter and substrate motion, is schematically shown in Fig. \ref{fig:sec_stage_illustration}. Since the spreading happens due to the vertical elongation of the droplet, which collapses, the increase in surface area during \textit{Stage-II} spreading ($\Delta SE_{II}$) can be approximated as the additional potential energy stored in the elongated droplet height, i.e.,  
\begin{equation}
   \Delta SE_{II}=m g \delta h= \frac{\pi}{6}\rho D_0^3 g V_s t'_{cap}.
\label{eq:dSE_II}
\end{equation}
Thus, the surface area and the diameter of the droplet after the \textit{Stage-II} spreading can be given by,
\begin{equation}
SE_{II}=(SE_f - \Delta SE_{rlx}) + \Delta SE_{II}=\gamma \pi \left( \frac{D_{max,II}^2}{4} + \frac23 \frac{D_0^3}{D_{max,II}}\right),
\label{eq:SE_II}
\end{equation}
where $(SE_f - \Delta SE_{rlx})$ is the surface area of the droplet before the \textit{Stage-II} spreading has initiated. Clearly, for $D_{max}$ to occur during \textit{Stage-II} spreading ($D_{max,II}>D_{max,I}$) 
the required condition is
\begin{equation}
\Delta SE_{II} > \Delta SE_{rlx}
\label{eq:dSE_cond}
\end{equation}
Furthermore, the aforementioned oscillation-assisted elongation of the droplet and subsequent \textit{Stage-II} spreading requires the downward motion of the substrate to initiate during the relaxation period (bounded by $\tau$ and $t_{jet}$) of the droplet. Hence the corresponding required condition is 
\begin{equation}
   \tau < T_{crest} < t_{jet}.
\label{eq:t_crest_cond}
\end{equation}
Here $T_{crest}$ is the time when the substrate initiates its downward motion, identified by the crest of the sinusoidal trajectory of the substrate motion (shown in Fig. \ref{fig:sec_stage_illustration}).
Equations \ref{eq:dSE_cond} and \ref{eq:t_crest_cond}, together, compose the necessary and sufficient condition for maximum spreading to occur in the \textit{Stage-II}.  
To test this scaling analysis for \textit{Stage-II}, in Fig. \ref{fig:D2minusD1vsSE}, we plotted the experimentally measured difference between the maximum spreading diameter of \textit{Stage-I} and \textit{Stage-II} as a function of $\Delta SE_{II}/\Delta SE_{rlx}$ for all experiments. We observe that all the data with $D_{max,II}>D_{max,I}$ (identified by green symbols) are located in a range of $\Delta SE_{II} > \Delta SE_{rlx}$. Similarly, the (red) data points which do not show the \textit{Stage-II} spreading ($D_{max,2}<D_{max,I}$), are mostly lying in the regime of $\Delta SE_{II}<\Delta SE_{rlx}$.  We also notice a few outliers, which show $D_{max,II}<D_{max,I}$, even though they satisfy the theoretical condition for \textit{Stage-II} spreading ($\Delta SE_{II} > \Delta SE_{rlx}$). It is to be noted for these points, the observed difference between $D_{max,I}$ and $D_{max,II}$ are rather small ($<5\%$), and most lie within the experimental uncertainty. Overall, we can infer that the experimental observation closely supports the necessary conditions derived for \textit{Stage-II} spreading.  

\begin{figure}
    \centering
    \includegraphics[width=1\linewidth]{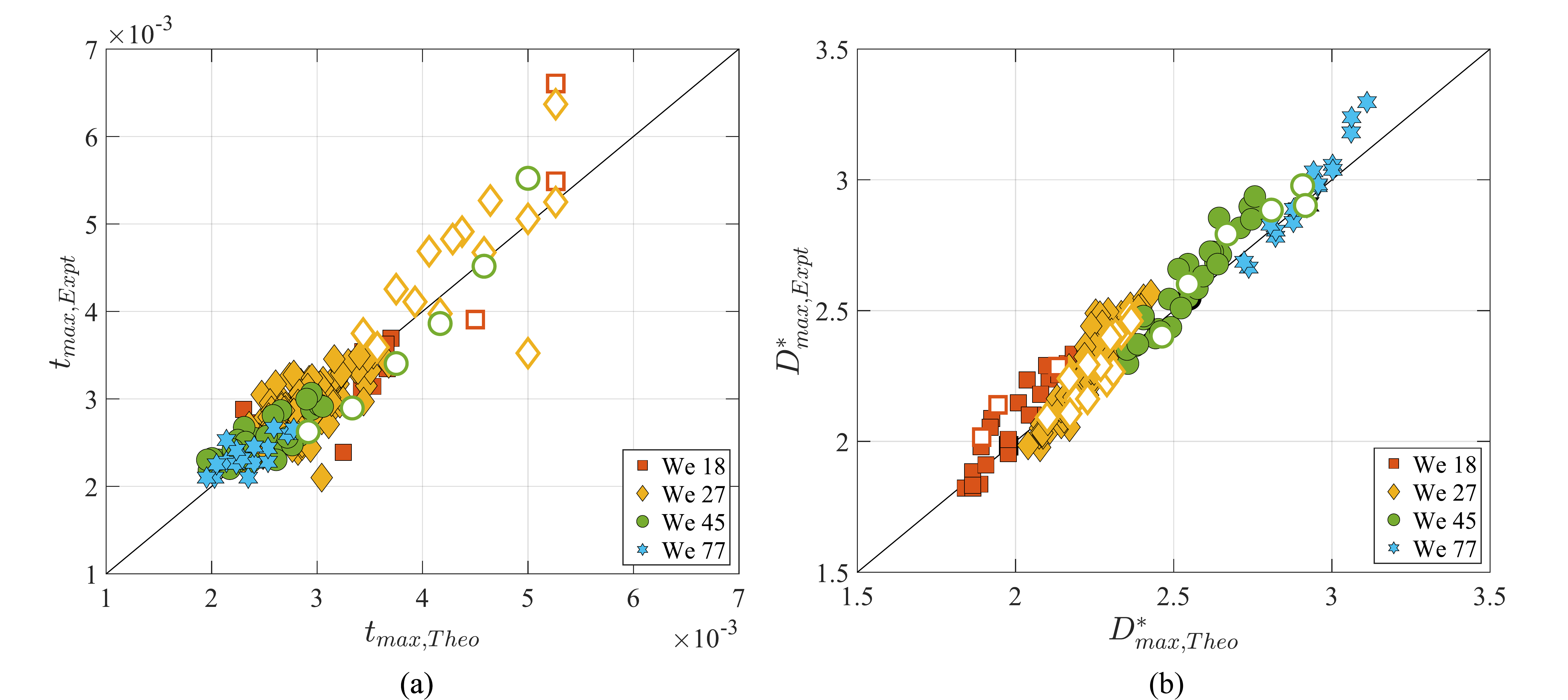}
\caption{Comparison of experimental and theoretically predicted values for $\textit{t}_{max}$ and $D^*_{max}$ at instant of maximum spreading taking theoretical estimate of $\textit{Stage-II}$ into consideration.}
    \label{fig:d_max_t_max_compare_all_Sec_Peak}
\end{figure}

Since \textit{Stage-II} spreading occurs at the instance when the downward motion of the substrate ends and the upward motion begins, the maximum spreading time can be approximated by $t_{max,II}\approx T_{trough}$. Finally, by using Eqs. \ref{eq:SE_relax} - \ref{eq:SE_II}, one can evaluate the theoretical approximation of $D_{max,II}$, the maximum diameter obtained in \textit{Stage-II}. In Fig. \ref{fig:d_max_t_max_compare_all_Sec_Peak} a and b, we compare the experimentally measured $t_{max}$ and $D^*_{max}$ with the theoretically obtained values for all experiments, including both \textit{Stage-I} and \textit{Stage-II} spreading. To evaluate the theoretical values, we used scaling for \textit{Stage-II} if the experimental condition satisfies Eq. \ref{eq:dSE_cond} and \ref{eq:t_crest_cond}. The comparison confirms that the presented scaling analyses can satisfactorily predict the spreading time and maximum spreading diameter.  

Before we conclude this section, we will discuss two additional observations, which can be made from Eq. \ref{eq:t_crest_cond}, one of the necessary conditions for \textit{Stage-II} spreading. From definitions, we know that $\tau \sim D_0/V_0$, $t_{jet}\sim \sqrt{{\rho D_0^3}/{\gamma}}$, and thus, $t_{jet}/\tau\sim \sqrt{We}$. Since Eq. \ref{eq:t_crest_cond} inherently implies $t_{jet}>\tau$, \textit{Stage-II} is expected to occur only for high impact inertia ($We>1$) conditions. Furthermore, by using the properties of a sinusoidal trajectory (See Fig. \ref{fig:sec_stage_illustration}), we can express $T_{crest}=(5\pi/4-\phi)/2\pi f$. Substituting this in Eq. \ref{eq:t_crest_cond} and recalling that $\phi>0$, we find $f>5/8t_{jet}$. This suggests that a minimum frequency is required for substrate oscillation to assist \textit{Stage-II} spreading. In experiments, indeed, \textit{Stage-II} spreading can be observed only for $f\geq250 Hz$. 

\section{Conclusion}
\label{sec:conclusion}
In summary, we presented a detailed experimental study highlighting the effects of an oscillating hydrophobic substrate on the spreading process of an impacting droplet. We showed that the maximum droplet diameter and the time taken are greatly affected by parameters of substrate oscillation, such as the frequency, amplitude, and phase at impact. The maximum spread can occur in two stages. The oscillation may promote or inhibit spreading in \textit{Stage-I}, which is primarily controlled by impact inertia. The scaling analyses showed that the instantaneous motion of the substrate alters the relative velocity between the droplet and the substrate, thereby changing the timescale of spreading. The energy budget confirms this change in spreading time will greatly affect the viscous dissipation; thus, the droplet attains a different spreading diameter. The phase of the oscillation at the impact greatly affects this process.

While inertia-controlled \textit{Stage-I} spreading is observed for all conditions, a secondary spreading process (\textit{Stage-II}) can be observed for certain impact conditions in which the droplet attains the maximum diameter at the latter stage. The \textit{Stage-II} spreading is controlled purely by substrate oscillations. For a higher frequency of oscillation, if the phase at the impact is such that the substrate initiates its downward motion after the \textit{Stage-I} spreading, the droplet undergoes sudden elongation. This allows the droplet to attain greater potential energy, and when it collapses, the droplet spreads to a larger diameter. Based on the scaling argument, the necessary conditions for the \textit{Stage-II} spreading were identified, which showed good agreement with experimental data. 

We end this exposition by discussing the relevance and limitations of the finding in the context of applications. In many industrial processes, droplet impact involves heat and mass transfer between the droplet and the substrate. Thus, it is of interest to modulate the spreading diameter and time. Our study provides a guideline to design the oscillation parameters for applications where substrate oscillations can be used for such modulation. It should be noted that the present study has been conducted in a frequency range of less than 1kHz, for which we expect the proposed scaling to hold. However, very different behavior of liquid droplets is observed for high (MHz-range) frequencies, where the current scaling may not be applicable.

\bibliographystyle{jfm}
\bibliography{References}

\end{document}